# Analysis of Droughts and Their Intensities in California from 2000 to 2020

Ujjwal[1], Shikha C. Patel[1], Bansari K. Shah[1], Nicholas Ogbonna[1], Huthaifa I Ashqar[1,2]

[1] Data Science Department, University of Maryland Baltimore County

[2] Arab American University, Jenin, Palestine

## Abstract

Drought has been perceived as a persistent threat globally and the complex mechanism of various factors contributing to its emergence makes it more troublesome to understand. Droughts and their severity trends have been a point of concern in the USA as well, since the economic impact of droughts has been substantial, especially in parts that contribute majorly to US agriculture. California is the biggest agricultural contributor to the United States with its share amounting up to 12% approximately for all of US agricultural produce. Although, according to a 20-year average, California ranks fifth on the list of the highest average percentage of drought-hit regions. Therefore, drought analysis and drought prediction are of crucial importance for California in order to mitigate the associated risks. However, the design of a consistent drought prediction model based on the dynamic relationship of the drought index remains a challenging task. In the present study, we trained a Voting Ensemble classifier utilizing a 'soft' voting system and three different Random Forest models, to predict the presence of drought and also its intensity. In this paper, initially, we have discussed the trends of droughts and their intensities in various California counties reviewed the correlation of meteorological indicators with drought intensities and used these meteorological indicators for drought prediction so as to evaluate their effectiveness as well as significance.

*Keywords: California, meteorological droughts, Random Forests, Voting Ensemble, Drought Severity, PDSI*

**Introduction**

Drought is a complex, and not an easily perceptible phenomenon that packs quite some disastrous consequences. As per Hao et al., "Drought is among the most disastrous natural hazards and occurs in virtually all geographical areas". (Hao, Singh, & Xia, 2018) Its silent occurrence makes it a more troublesome threat as compared to other natural disasters like a tornado or a hurricane. "It's often described as a "creeping phenomenon" because it slowly impacts many sectors of the economy and operates on many different timescales". ((NIDIS))

According to the National Integrated Drought Information System or NIDIS, "Drought is generally defined as "a deficiency of precipitation over an extended period of time (usually a season or more), resulting in a water shortage" ". ((NIDIS)) But, there are a number of complex factors in play that gives rise to drought in a region. Hao et al. describe that "Overall, the development and evolution of drought result from complicated interactions among meteorological anomalies, land surface processes, and human activities". (Hao, Singh, & Xia, 2018)

**Figure 1:**

Image Source: *"Climate-influenced relationships between hydrological cycle variables and drought types"* (Kim & Jehanzaib, 2020)

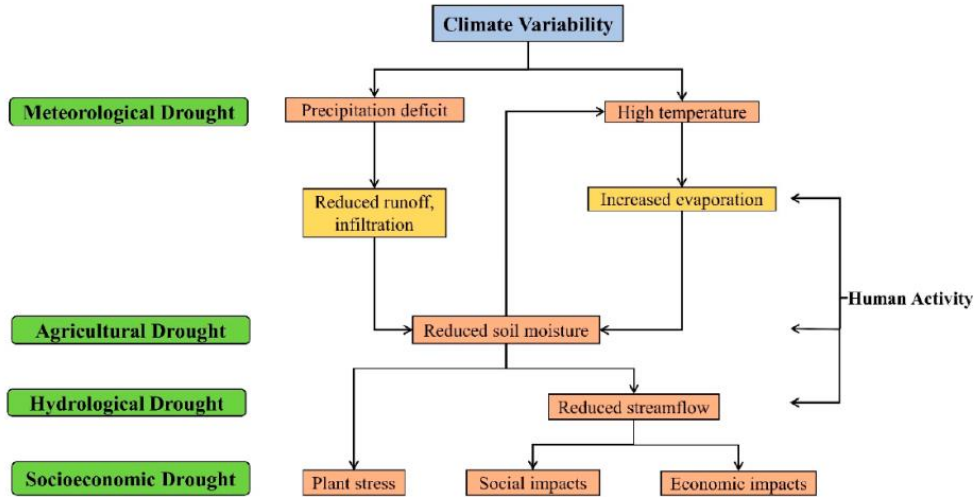

Droughts are generally categorized into, "meteorological, agricultural, hydrological, and socioeconomic drought, based on both physical and socioeconomic factors". (Hao, Singh, & Xia, 2018) In general, the other types of droughts originate as a consequence of the onset of meteorological droughts. "Therefore, the prediction of meteorological droughts is the most important for the mitigation of cascaded impacts of different kinds of droughts". (Malik, Tikhamarine, Sammen, Abba, & Shahid, 2021)

**Table 1:**

*Drought types and their descriptions*

| Drought Type | Description |
|---|---|
| **Meteorological Drought** | Precipitation deficit over a prolonged period of time |
| **Agricultural Drought** | Deficit in soil moisture, which affects plant production and crop yield. |
| **Hydrological Drought** | Deficit of surface runoff, streamflow, reservoir, or groundwater level. |
| **Socioeconomic Drought** | associated with the supply and demand of some economic goods (e.g., water, food grains), which incorporates features or impacts of the other three types of drought |

Preventing a drought is beyond human capabilities and therefore risks mitigation and management remains our only option of dealing with this hazard. Due to this fact, the presence of a reliable drought prediction and forecasting system is crucial. "The ultimate objective of drought prediction is to prepare a mitigation plan in advance, rather than resolve intellectual curiosity about nature. Drought forecasting plays an important role in mitigating the negative effects of drought". (Park, Kim, & Lee, 2019) The dangers associated with droughts are being taken seriously all around the world. "Drought risk analysis, forecasting and assessment are facing rapid expansion, not only from theoretical but also practical points of view. Accurate monitoring, forecasting and comprehensive assessments are of the utmost importance for reliable drought-related decision-making". (Kim & Jehanzaib, 2020)

**Literature Review**

"Drought is among the most damaging and least understood of all natural hazards, which causes tremendous economic and social impacts". (Huang, et al., 2014) Several regions of the USA have been hit with droughts time and again, which have led to substantial economic losses. "Since 2000, 12 drought events have each caused greater than one billion dollars in economic impacts, with some events costing the country over $10 billion". (Huang, et al., 2014) "California (CA) experienced an exceptional drought (2011–17) causing severe economic and ecological losses—including $5.5 billion in agricultural losses from 2014 through 2016". (Kam, Stowers, & Kim, 2019) Therefore, the drought research community are making rigorous efforts to analyse the factors responsible for the drought.

"Drought prediction is a major concern for water managers, farmers, and other final users because it limits their decisions". (Aghelpour & Varshavian, 2021) "Accurate forecasting of drought is therefore essential for multiple water resources planning, optimal operation of the irrigation system, drought preparedness, and mitigation". (Malik A, 2020) Drought prediction of meteorological droughts is more important in this regards since, "All kinds of droughts are initiated from meteorological droughts". (Malik, Tikhamarine, Sammen, Abba, & Shahid, 2021)

"Drought prediction generally refers to the prediction of drought severity (e.g., values of a specific drought indicator)". (Hao, Singh, & Xia, 2018) There are a number of standard indices used to measure the severity of meteorological droughts. "Meteorological droughts are generally quantified using different indices like Palmer Drought Severity Index (PDSI), Reconnaissance Drought Index (RDI), Standardized Precipitation Index (SPI), Standardized Precipitation, Evapotranspiration Index (SPEI), and Effective Drought Index (EDI)".

To better prepare for and mitigate drought impacts, we are encouraged to work on current drought patterns, their impacts and their evolution in specific regions.

We focused on addressing four research questions in this paper. First, we intended to analyse the patterns of occurrence of different intensity droughts in all California counties from 2000 to 2020. Examining this aspect helped us in understanding how the frequency of each type of drought varied over the last 21 years. Subsequently, we were able to figure out which kind of drought dominated in each of these years. Second, we wanted to understand how strong the connection between drought in a region and its meteorological data was. Third, we wished to investigate how accurately we can predict whether a region is drought-hit or not as well as the intensity of drought-hit region, with the help of meteorological indicators. It helped us in obtaining a fairly confident model that could prove advantageous as an early warning system. Lastly, we aimed at finding out which of these indicators were the most significant in predicting droughts and their intensities. This will be insightful in conducting research on developing instruments that capture those features more accurately.

**Datasets**

We have used the datasets available on 'Kaggle' with the name 'Predict Droughts using Weather & Soil Data', which has been uploaded by Christoph Minixhofer. The data owes its origin to the NASA Langley Research Centre (LaRC) POWER Project funded through the NASA Earth Science/Applied Science Program.

Originally, there were four datasets namely 'test_timeseries', 'train_timeseries', 'validation_timeseries' and 'soil_data'. Each of 'test_timeseries', 'train_timseries' and 'validation_timeseries' consists of 18 meteorological features as well as a score label. The values of all 18 meteorological features are available for each day and the values of the score label are available for each week. The 'train_timeseries' consists of data from 2000 to 2016, the 'validation_timeseries' consists of data from 2017 to 2018 and the 'test_timeseries' consists of data from 2019 to 2020.

**Table 2:**

*18 meteorological features and their descriptions*

| Meteorological Feature | Description |
|---|---|
| PRECTOT | Precipitation (mm day$^{-1}$) |
| PS | Surface Pressure (kPa) |
| QV2M | Specific Humidity at 2 Meters (g/kg) |
| TS | Earth Skin Temperature (C) |
| WS50M | Wind Speed at 50 Meters (m/s) |
| WS50M_MIN | Minimum Wind Speed at 50 Meters (m/s) |
| WS50M_MAX | Maximum Wind Speed at 50 Meters (m/s) |
| WS50M_RANGE | Wind Speed Range at 50 Meters (m/s) |
| WS10M | Wind Speed at 10 Meters (m/s) |
| WS10M_MIN | Minimum Wind Speed at 10 Meters (m/s) |
| WS10M_MAX | Maximum Wind Speed at 10 Meters (m/s) |
| WS10M_RANGE | Wind Speed Range at 10 Meters (m/s) |
| T2M | Temperature at 2 Meters (C) |
| T2M_MIN | Minimum Temperature at 2 Meters (C) |
| T2M_MAX | Maximum Temperature at 2 Meters (C) |
| T2M_RANGE | Temperature Range at 2 Meters (C) |
| T2MDEW | Dew/Frost Point at 2 Meters (C) |
| T2MWET | Wet Bulb Temperature at 2 Meters (C) |

The 'score' label value indicates the severity of the drought. These values are based on Palmer Drought Severity Index (PDSI) values which is one of the standard score used for measuring drought intensity. "PDSI is used to define and monitor drought which attempts to measure the duration and intensity of long-term, spatially extensive drought, based on precipitation, temperature, and available water content data." (Arthur & Saffer) These values are continuous values ranging from 0 to 5 and can be classified into 5 drought intensity indicators (D0, D1, D2, D3 and D4), representing the magnitude of the drought.

**Table 3:**

*Drought Intensity Labels and their Descriptions*

| Label | Description |
|---|---|
| D0 | Abnormally Dry |
| D1 | Moderate Drought |
| D2 | Severe Drought |
| D3 | Extreme Drought |
| D4 | Exceptional Drought |

The 'soil_data' consists of 'fips' (US County FIPS code), 'Latitude', 'Longitude' and 29 soil related features. This dataset owes its origin to Harmonized World Soil Database hosted by 'Food and Agricultural Organization of the United Nations'.

Additionally, we retrieved data for County FIPS codes from the website of 'United States Department of Agriculture' in order to get a dataset with each county's name, state and FIPS code. FIPS stands for Federal Information Processing Standard Publication 6-4 which is a five digit code to uniquely identify each US County. We created a dataset named 'fips' which consisted of FIPS code of the US county, its name and the State it belongs to; denoted by the FIPS, Name and State respectively.

## Methodology

**Data Preparation**

Our study is only concerned with California as it is one of the biggest agricultural contributors in the U.S.A. and it experiences droughts frequently. From 2000 to 2020, California's "20-year average places it at number 5 on the list, with 59.54% of the state in drought during the winter and 63.40% in the fall". (Drought Risk in the United States and State-Specific Insights, 2020).

First, we combined 'train_timeseries', 'validation_timeseries' and 'test_timeseries' into one single dataset named 'full_dataset_sorted' which then consisted of all the meteorological features and drought intensity data from 2000 to 2020. This dataset consisted of the data for all U.S. Counties. Using the 'fips' dataset, we filtered the 'full_dataset_sorted' with the help of California counties FIPS codes and created another dataset named 'full_dataset_CA_sorted'.

As mentioned earlier, 18 meteorological features were recorded daily and 1 drought intensity label was recorded on a weekly basis, making it difficult to analyse the data varying on two different time scales. To eliminate this problem, we aggregated the last 90 days of meteorological data and combined it with the corresponding drought intensity reading to prepare the final dataset named 'ninety_days_aggregated_data_ca_drought'. This final dataset had 63568 rows and 19 columns, and we used it for analysing the trends of droughts and training the model for drought prediction.

**Model training**

We divided our model training tasks into two subtasks. First, we predicted whether reading in the dataset indicates the presence or absence of drought. And second, we predicted the intensity of the drought in case it is present.

Since the range of our meteorological feature values varied substantially, we had to apply the Min-Max Normalization technique to normalize the dataset to a common scale i.e. all the meteorological feature values were rescaled in the range of [0,1].

For our first subtask, the dataset was split into 70% and 30% for the training data and testing data respectively. Initially, we trained a Support Vector Machine (SVM) model, Logistic Regression model, a K-Nearest Neighbour (KNN) model, an XGBoost model, and a Random Forest model in order to understand which of these models gives the best results. For our dataset, Random Forest performed the best and hence we created three different variants of Random Forests and fed them to a Voting Ensemble classifier which used the 'soft' voting method.

Similarly, for our second subtask, the dataset was split into 70% and 30% for the training data and testing data respectively. Again, we trained a Support Vector Machine (SVM) model, Logistic Regression model, a K-Nearest Neighbour (KNN) model, an XGBoost model, and a Random Forest model in order to understand which of these models gives the best results. Since our second subtask was a multi-label classification problem, we deployed those models using OneVsRest Classifier. For our dataset, Random Forest again performed the best. Hence, we created three different variants of Random Forests and fed them to a Voting Ensemble classifier which used the 'soft' voting method.

**Figure 2:**

*Methodology flow chart*

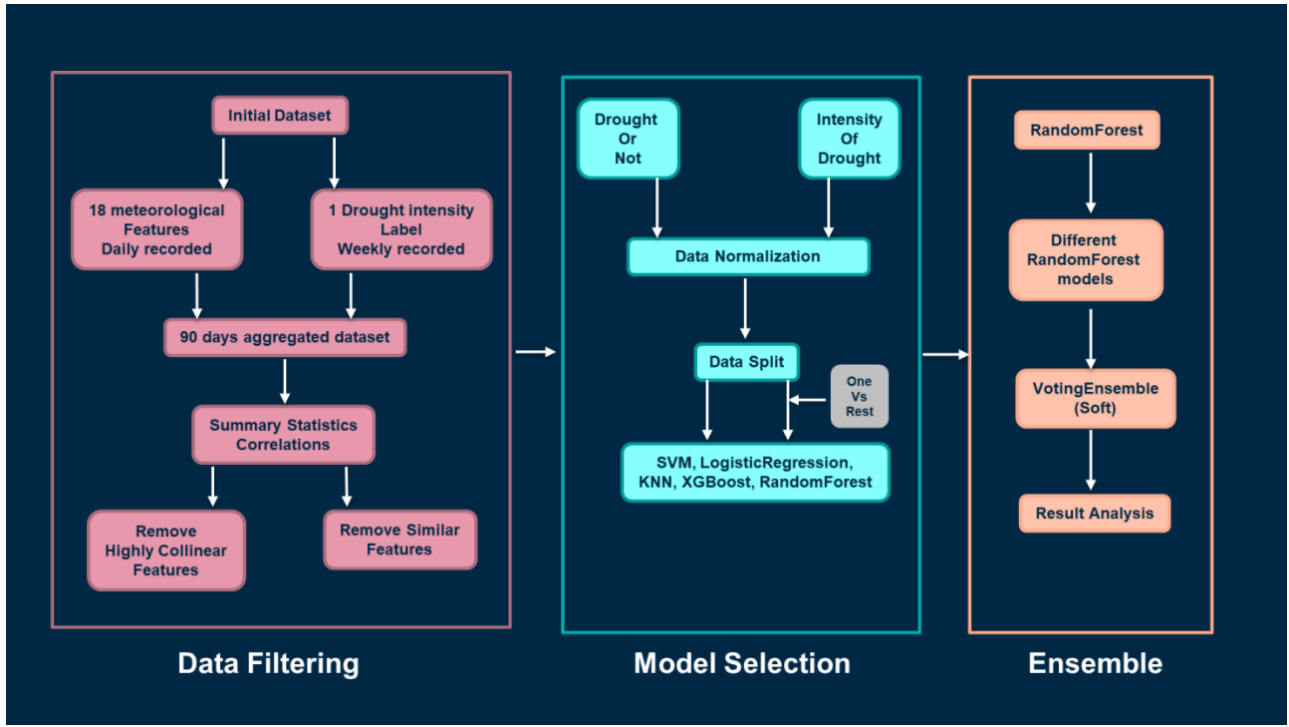

## Analysis

**Study of frequencies of different intensity droughts from the year 2000-2020:**

**Figure 3:**

*Frequency of each type of drought every year from 2000 to 2020*

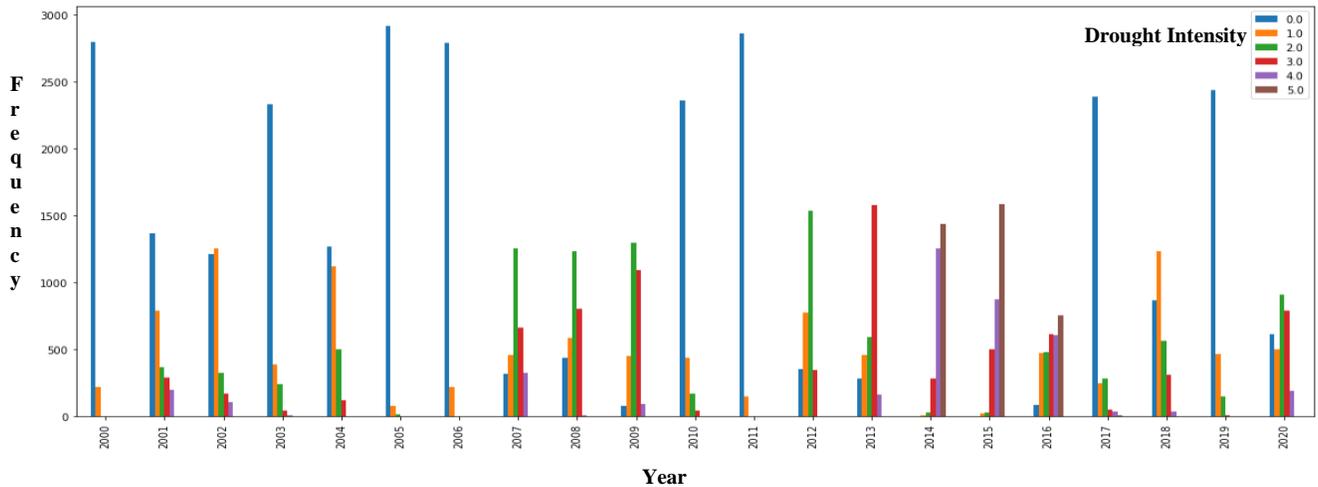

The bar graph shown below represents the number of occurrences of each 'score' intensity label for every year from 2000 to 2020, for all the California Counties combined. These labels include, '0' that means 'no drought' conditions as well as each type of drought intensity label i.e. drought with an intensity level of D0 (Abnormally Dry), D1 (Moderate Drought), D2 (Severe Drought), D3 (Extreme Drought) and D4 (Exceptional Drought). The y-axis shows the number of times each intensity label appeared in the dataset for a given year, and the x-axis displays the year. '0' is denoted by a blue bar which represents the absence of drought. 'D0', 'D1', 'D2', 'D3' and 'D4' are represented by orange, green, red, purple and brown bars respectively. The sudden emergence of extreme drought (D3) and exceptional drought (D4) are observed from 2014 onwards. Also, from 2014 to 2020, a notable decrease is evident in occurrences of '0' labels. Furthermore, no occurrence of exceptional drought (D4) was observed until 2014. As for the year 2020, the majority of the occurrences consists of extreme drought (D3) and exceptional drought (D4) and therefore it can be deduced that drought will continue to remain a persistent threat for California in the near future.

The 6 maps displayed below, highlights the state of California along with its 58 Counties. The bubbles are placed according to the location of each California County and represent the difference in the percentage of occurrences for a particular type of drought label. The frequency for each of the drought intensity labels was retrieved as a percentage value for the years 2000-2013, 2007-2013, and 2014-2020. Then, the differences in these percentages for each California County were calculated for 2000-2013 and 2014-2020 (first scenario) as well as for 2007-2013 and 2014-2020 (second scenario). The following figure describes the comparison of the percentage of occurrence of each label for 2000-2013 and 2014-2020. The number of Counties experiencing positive change in percentage and negative change in percentage was analysed. On the right of each of the map, the colour scale present depicts the mapping of the colour of the bubble to a value (positive or negative). The size of the bubble illustrates the magnitude of change.

**Figure 4:**

*Change in percentage of occurrences of each drought intensity label for the year 2000-2013 and 2014-2020*

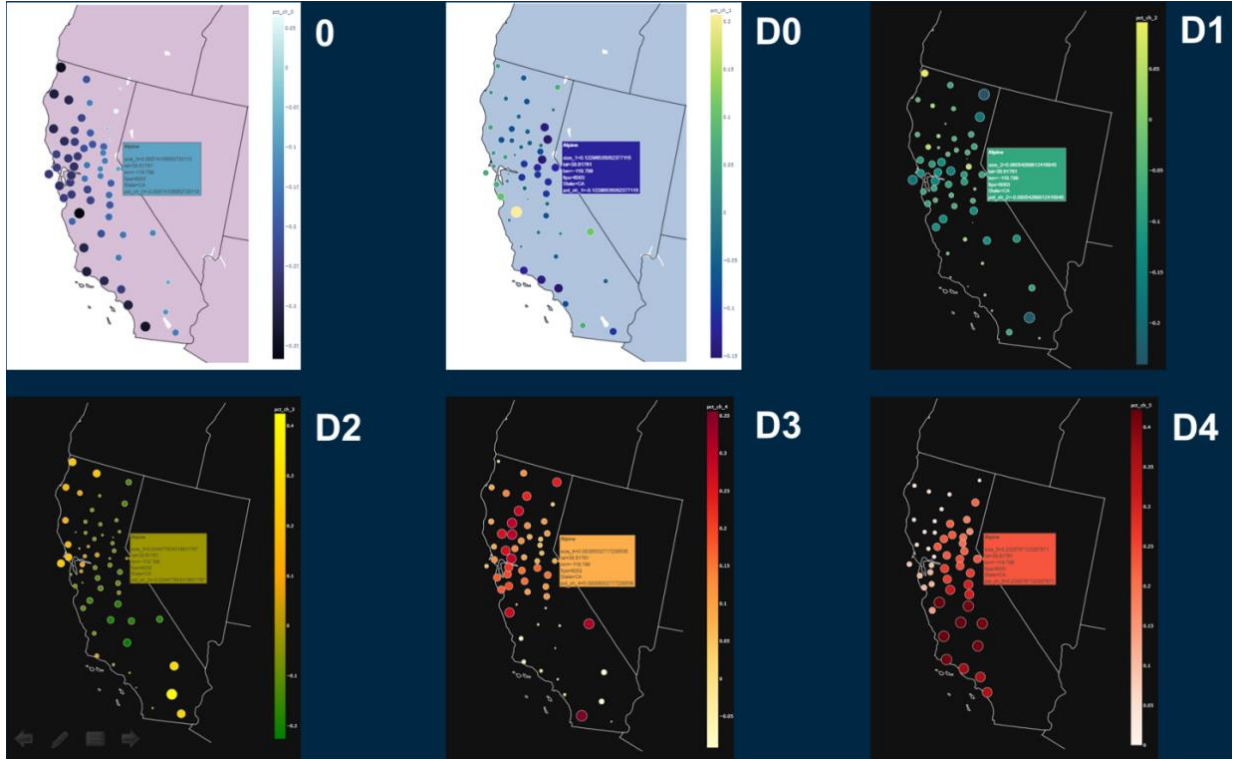

### *'0' or 'No Drought' Labels Trend*

For the first scenario, 'No Drought' conditions resulted in 3 counties experiencing positive change and 55 counties witnessing negative change. This means that the percentage of drought occurrences was higher from 2014 to 2020 than from 2000 to 2013. For the second scenario, 57 counties revealed a positive change and only one county showcased a negative change, implying that there has been less drought occurrence in 2014-2020 as compared to 2007-2013. However, the actual numbers for this scenario show that the percentage change was quite small.

### *'D0' or 'Abnormally Dry' Trend*

For the first scenario, 16 counties displayed positive change and 42 counties encountered negative change. This means that the percentage of 'D0' drought occurrences was higher from

14
CALIFORNIA DROUGHT ANALYSIS2014 to 2020 than from 2000 to 2013. For the second scenario, 48 counties showed a positive change, and 10 counties exhibited a negative change, implying that there have been fewer occurrences of 'D0' droughts in 2014-2020 than in 2007-2013.

### *'D1' or 'Moderate Drought' Trend*

For 'D1' labels, 10 counties saw a positive change, while 48 counties saw a negative change in the first scenario. Likewise, 19 counties indicated a positive change while 39 counties indicated a negative change in the second scenario.

### *'D2' or 'Severe Drought' Trend*

For the first scenario, 22 counties experienced positive change and 36 counties endured negative change for 'D2'. For the second scenario, 25 counties indicated a positive change, while 33 counties showed a negative change.

### *'D3' or 'Extreme Drought' Trend*

For the first scenario, 50 counties exhibited positive change and 8 counties exhibited negative change for 'D3' labels. As for the second scenario, 52 counties showed a positive change, while 6 counties showed a negative change. This implies that there have been more occurrences of 'D3' droughts in 2014-2020 as compared to 2000-2013 and 2007-2013.

### *'D4' or 'Exceptional Drought' Trend*

As evident in Figure 1 and as mentioned earlier, there has not been a single occurrence of the 'D4' label before 2014. Therefore, all of the 58 counties experienced positive change for 'D4' labels clearly implying that the percentage of 'D4' drought occurrences was higher from 2014 to 2020 as compared to 2000-2013 as well as 2007-2014.

## Results

**Drought Prediction**

The three different Random Forest models created by varying 'n_estimators' parameter, predicted with an accuracy score of 0.84657, 0.84631 and 0.84327. The Voting Ensemble Classifier employed a 'soft' voting method on these three models and performed with an accuracy of 0.84642.

**Figure 4:**

*Prediction Results for presence of drought*

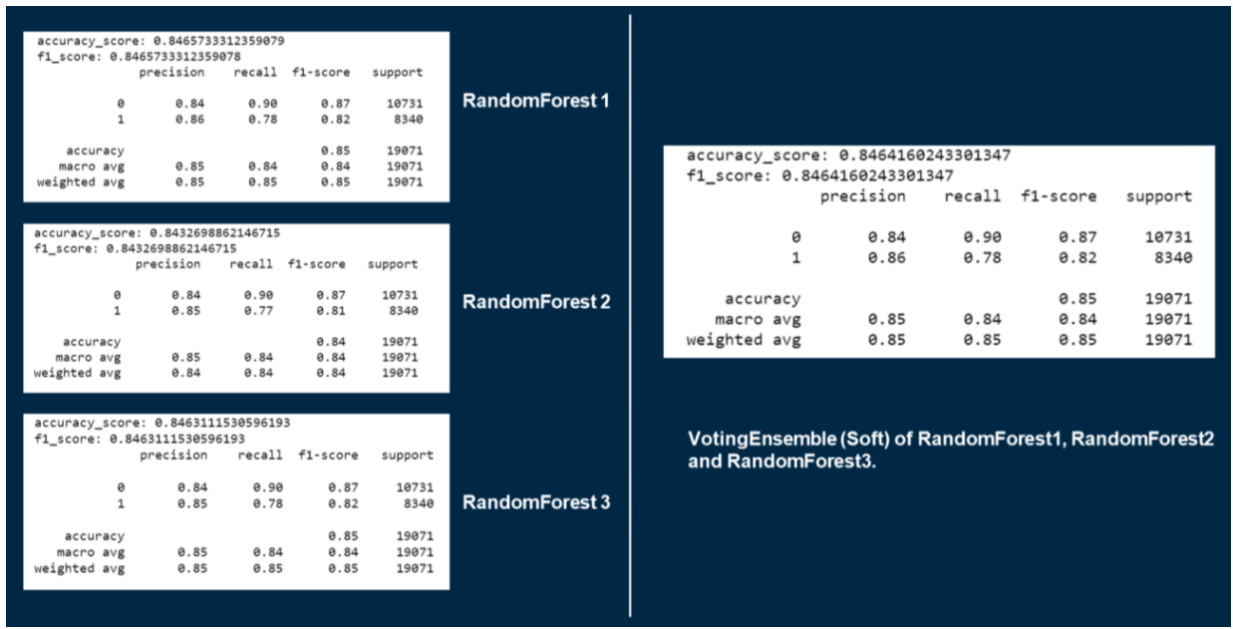

**Drought Intensity Prediction**

The three different Random Forest models created by varying 'n_estimators' parameter, coupled with OneVsRest Classifier produced an accuracy score of 0.72240, 0.73125 and 0.73218. The Voting Ensemble Classifier utilizing these three Random Forest models along with the 'soft' voting method, performed with an accuracy of 0.73367.

**Figure 5:**

*Prediction results for drought intensity*

```
accuracy_score: 0.7224033535165347
f1_score: 0.7224033535165348
              precision    recall  f1-score   support

           1       0.72      0.77      0.75      3126
           2       0.67      0.75      0.71      2970
           3       0.71      0.71      0.71      2323
           4       0.81      0.57      0.67      1164
           5       0.84      0.71      0.77      1152

    accuracy                           0.72     10735
   macro avg       0.75      0.70      0.72     10735
weighted avg       0.73      0.72      0.72     10735
```
RandomForest 1

```
accuracy_score: 0.7321844434094085
f1_score: 0.7321844434094085
              precision    recall  f1-score   support

           1       0.73      0.78      0.75      3126
           2       0.68      0.76      0.72      2970
           3       0.72      0.72      0.72      2323
           4       0.82      0.59      0.69      1164
           5       0.86      0.71      0.78      1152

    accuracy                           0.73     10735
   macro avg       0.76      0.71      0.73     10735
weighted avg       0.74      0.73      0.73     10735
```
RandomForest 2

```
accuracy_score: 0.7312529110386586
f1_score: 0.7312529110386586
              precision    recall  f1-score   support

           1       0.73      0.78      0.75      3126
           2       0.68      0.75      0.72      2970
           3       0.72      0.72      0.72      2323
           4       0.82      0.59      0.69      1164
           5       0.86      0.71      0.78      1152

    accuracy                           0.73     10735
   macro avg       0.76      0.71      0.73     10735
weighted avg       0.74      0.73      0.73     10735
```
RandomForest 3

```
accuracy_score: 0.7336748952026083
f1_score: 0.7336748952026082
              precision    recall  f1-score   support

           1       0.73      0.78      0.76      3126
           2       0.68      0.76      0.71      2970
           3       0.73      0.72      0.73      2323
           4       0.83      0.60      0.69      1164
           5       0.86      0.70      0.77      1152

    accuracy                           0.73     10735
   macro avg       0.77      0.71      0.73     10735
weighted avg       0.74      0.73      0.73     10735
```
VotingEnsemble (Soft) of RandomForest1, RandomForest2 and RandomForest3.

**Most Significant Features**

We analysed the significance of meteorological features in 3 scenarios. In the first scenario, we fed the dataset to a Random Forest model without removing any features which predicted with an accuracy of 0.84657. In the second scenario, we removed highly collinear features and got an accuracy score of 0.8527 and in the third scenario; we removed similar features like T2M, T2M_RANGE, T2M_MIN etc. and got an accuracy of 0.8155. In all three models, Precipitation (PRECTOT), Surface Pressure (PS) and Specific Humidity at 2 Meters (QV2M) were found to be the most significant features.

**Figure 6:**

*Feature Significance in Different Models*

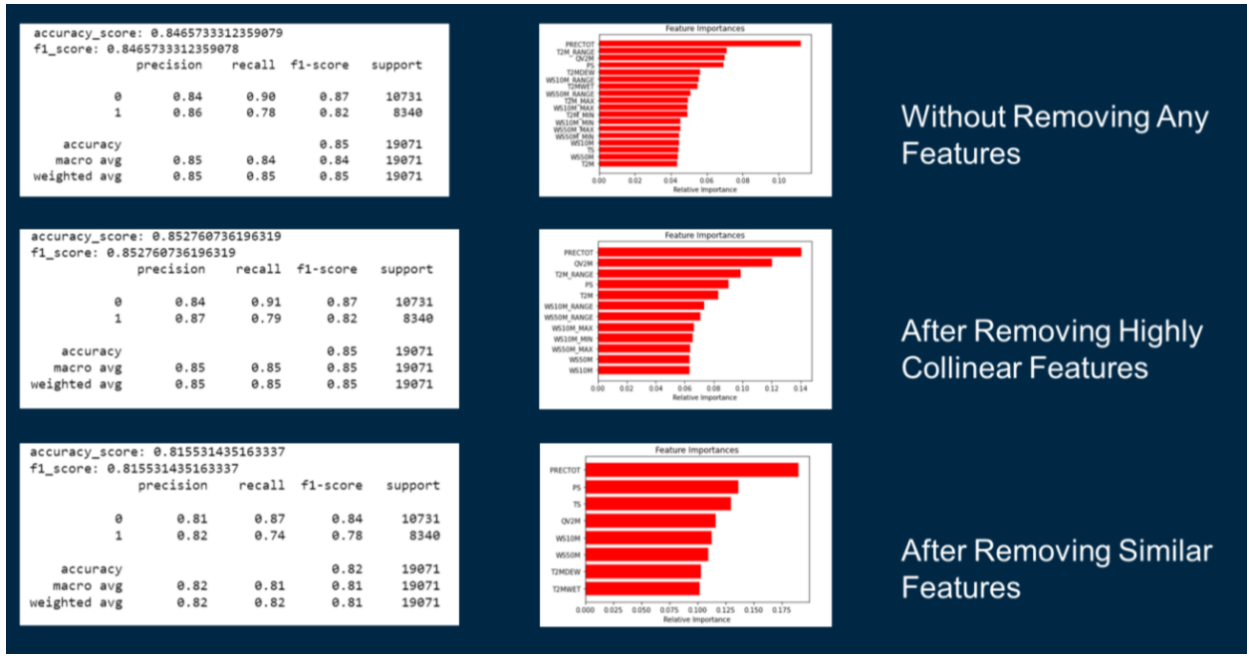

**Conclusions**

From our research, we can conclude that there has been a substantial increase in the severity of droughts in California from 2014 onwards. Judging from the last 7 years data, it is highly likely that the intense droughts will keep on persisting in the coming years. Since preventing droughts is beyond our capabilities, there is a need for an advanced risk management system to counter the effects of droughts. This necessitates the demand for a reliable and robust drought prediction system. From our study and model training results, it is evident that meteorological features are significantly helpful in predicting the presence of drought and also its intensity. Furthermore, it has been observed that precipitation, surface pressure and specific humidity are crucial in prediction task and therefore more emphasis should be given on capturing data related to them and developing technologies to accurately capture these features. Finally, Random Forests coupled with Voting Ensemble techniques can be used as a potentially helpful

18
CALIFORNIA DROUGHT ANALYSISmodel for drought prediction tasks. It can be adopted as an early warning system to sense the presence of drought in a region and hence further research can be conducted on this topic.

## Future Scope of Study

In our research, the drought intensities for California's various counties were predicted using 90-days aggregated data of meteorological features. The scope of the project can be expanded further by varying this time frame, for example, from 30 days to 180 days. Moreover, the location data can also be incorporated to make the models more confident. Also, the integration of time series trends like cyclicity and seasonality can be explored. The scope of this paper can also be expanded to all US counties or the ones that contribute significantly to US agriculture. For this study, we have considered only the Palmer Drought Severity Index (PDSI) score values for analysing and predicting the drought intensities. The scope of prediction can be further widened by considering other drought severity indexes like Integrated Drought Index (IDI), Standardized Precipitation Index (SPI), Palmer Crop Moisture Index, Palmer Z Index and Palmer Hydrological Drought Index.

## Acknowledgements

We would like to express our heartfelt gratitude to Dr. Huthaifa Ashqar for his relentless encouragement, assistance, and guidance, which was essential in the successful completion of this project. We would also like to thank Kaggle and Christoph Minixhofer for providing us with the complete dataset of meteorological indicators and drought severity labels. Further, we are grateful to NASA POWER Project, NASA Earth Science/Applied Science Program and the authors of the US Drought Monitor, without whom the availability of this dataset for open use would not have been possible. Finally, we are thankful for the Harmonized World Soil Database by Fischer, G., F. Nachtergaele, S. Prieler, H.T. van Velthuizen, L. Verelst and D. Wiberg.